\documentclass[12pt]{article}

\usepackage{amsmath,amsthm,amssymb}

\newcommand{\antich}{S}
\newcommand{\antichs}{\mathcal{S}}
\newcommand{\causet}{\mathcal{P}}
\newcommand{\chn}{C}
\newcommand{\chns}{\mathcal{C}}
\newcommand{\dff}{\sc}

\newcommand{\ie}{\emph{i.e.}\,\,}
\newcommand{\events}{\mathcal{P}}
\newcommand{\foliation}{\mathcal{F}}

\newcommand{\histories}{\mathcal{C}}
\newcommand{\history}{{C}}

\newcommand{\eg}{\emph{e.g.},\,}

\newtheorem{theo}{Theorem}
\newtheorem{conj}{Conjecture}

\title{Spacetime topology from the tomographic histories
approach: Part II}

\author{Ioannis Raptis\footnote{EU Marie Curie Reintegration Postdoctoral Research Fellow,
Algebra and Geometry Section, Department of Mathematics,
University of Athens, Panepistimioupolis, Athens 157 84, Greece;
{\em and} Visiting Researcher, Theoretical Physics Group, Blackett
Laboratory, Imperial College of Science, Technology and Medicine,
Prince Consort Road, South Kensington, London SW7 2BZ, UK; e-mail:
i.raptis@ic.ac.uk}, Petros Wallden\footnote{Theoretical Physics
Group, Physics Department, Imperial College; email:
petros.wallden@imperial.ac.uk}  and Rom\`an R.
Zapatrin\footnote{Department of Information Science, The State
Russian Museum, Inzenernaya 4, 191186, St.Petersburg, Russia;
email: zapatrin@rusmuseum.ru}}

\begin{document}

 \maketitle
 \begin{abstract}
 As an inverse
problem,  we recover the topology of the effective spacetime that
a system lies in, in an operational way. This means that from a
series of experiments we get a set of points corresponding to
events. This continues the previous work done by the authors. Here
we use the existence of upper bound in the speed of transfer of
matter and information to induce a partial order on the set of
events. While the actual partial order is not known in our
operational set up, the grouping of events to (unordered) subsets
corresponding to possible histories, is given. From this we
recover the partial order up to certain ambiguities that are then
classified. Finally two different ways to recover the topology are
sketched and their interpretation is discussed.
 \end{abstract}

 \section*{Introduction}\label{intro}

 The work presented here, is a continuation of \cite{case I} in
 which we introduced the concepts of inverse histories, effective
 topology and operationalistic means of recovering the background
 structure of effective spacetime. In that paper we dealt with the
 non-relativistic case, while here we proceed to the relativistic
 one.We should stress here, that relativistic is understood in the
 sense that there exists an upper bound in the speed of transfer
 of matter and information and the latter induces a partial (causal)
 order on the set of events. This leads
to further restrictions on the set of possible
histories/`trajectories', that in its turn, results to
 some proximity relation from the bare set of possible histories.

\paragraph{Motivation.} We want to recover in an operational way
the `arena' that the under consideration system, lies in. The
spacetime/arena that we will be talking, will be an effective one,
since we refer to spacetime ONLY according to the information we
have from our data, i.e. from some collection of local
information\footnote{see for works in a similar line of thought
\cite{Kribs:2005yc,Lloyd:2005js}}.

A record of a trajectory, for example in a cloud chamber
experiment (if the initial state is unknown) does not give us
information about the direction the particle went. We can only
deduce the order (up to an overall flip) because we assume certain
things about the dynamics of the system and assume that the
trajectory is continuous (and smooth in the metric of the given a
priori background structure). In our case the background is not
given and due to the restrictions of causality and having the set
of possible alternative histories, we will deduce which events are
close and therefore deduce the (causal) structure of the effective
spacetime.

This relate to the general philosophical issue of how we
understand from a frozen picture (final time record) the notion of
change in the past (see also Barbour \eg \cite{EOT}). In
particular in our operationalistic approach, we do not have access
to any other information but the record, which as a frozen
picture, does not have information about the order if no other
assumptions about the dynamics are made. The whole thing would
correspond in having a set of successive pictures each of which
corresponds to a different possible trajectory (and NOT of
different events, since each record is a whole history). Clearly,
in each picture we would know the points the system crossed
(`sub-records'), but the order that it crossed them, would be
unknown. In the photograph, the metric of the Euclidean space of
the `photographic plate' would give us already the order up to
total direction flip, of the events which in this case would
correspond to points the system visited. In our case we would have
two complications. First there is no background structure to
deduce which point is close spatially to which and therefore by
continuity and smoothness of the trajectory to deduce the order of
the events, and second we have more complex causal curves, i.e.
not order but partially ordered set . We need to get some notion
of proximity on these events based only on the records. We will be
describing two ways to do this in the paper, the one using the
variation of records (see below) and the other using the partial
order and the set of all possible different trajectories.

We now proceed to remind few concept from \cite{case I}.

 \paragraph{Histories and Inverse Histories.}
The \emph{standard decoherent} histories approach to quantum
mechanics deals with the kind of questions that may be asked about
a closed system, without the assumption of wavefunction collapse
(upon measurement). It tells us, in a non-instrumentalist way,
under what conditions we may meaningfully talk about statements
concerning histories of our system, by using ordinary logic. This
approach was mainly developed by Gell-Mann and Hartle
\cite{GH90a,GH90b,GH90c,Har91a,Har91b,GH92,Har93a}, and it was
largely inspired by the original work of Griffiths \cite{Gri84}
and Omn\`es \cite{Omn88a,Omn88b,Omn88c,Omn89,Omn90,Omn92}. A more
mathematically sophisticated formulation is due to Isham and
collaborators in \eg \cite{Isham:1993qs} \footnote{H.P.O. History
Projection Operator.}. This formulation, consists of a space of
histories $\mathcal{UP}$, which is the space of all possible
histories of the closed system in question, and a space of
decoherence functionals $\mathcal{D}$. Parenthetically, the space
of histories is usually assumed to be a tensor product of copies
of the standard Quantum Mechanics' Hilbert space. Two histories
are called disjoint, write $\alpha\perp\beta$, if the realization
of the one excludes the other. Two disjoint histories can be
combined to form a third one $\gamma=\alpha\vee\beta$ (for
$\alpha\perp\beta$). A complete set of histories is a set
$\{\alpha_i\}$ such that
$\alpha_i\perp\alpha_j\quad(\forall\alpha_i,\alpha_j,\quad i\neq
j)$, and
$\alpha_1\vee\alpha_2\vee\ldots\vee\alpha_i\ldots=\mathbf{1}$

A decoherence functional is a complex valued function
$d:\mathcal{UP}\times\mathcal{UP}\longrightarrow\mathbb{C}$ with
the following properties:
\begin{itemize}\item[a)]Hermiticity: $d(\alpha,\beta)=d^*(\beta,\alpha)$
\item[b)]Normalization: $d(1,1)=1$ \item[c)]Positivity:
$d(\alpha,\alpha)\geq0$ \item[d)]Additivity:
$d(\alpha,\beta\oplus\gamma)= d(\alpha,\beta)+d(\alpha,\gamma)$
for any $\beta\perp\gamma$

\end{itemize}

A complete set of histories $\{\alpha_i\}$ is said to obey the
{\dff decoherence} condition, \ie
$d(\alpha_i,\alpha_j)=\delta_{ij}p\left(\alpha_i\right)$ while
$p\left(\alpha_i\right)$ is interpreted as the probability for
that history to occur \emph{within the context of this complete
set}. The decoherence functional encodes the initial condition as
well as the evolution of the system. Here we should also note that
the topology of the space-time is presupposed when we group
histories into complete sets, \ie in collections of partitions of
unity.

In Quantum Mechanics, histories correspond to time ordered strings
of projections and to combination of these when they are disjoint.
An important issue here is the relation between decoherence and
records. Namely, it can be shown that if a set of histories
decoheres, there exists a set of projection operators on the final
time that are perfectly correlated with these histories and vice
versa.\footnote{This is the case for a \emph{pure} initial state,
and we restrict ourselves to it.} These projections are called
{\em records}. It is this concept that figures mainly in our
approach (\eg see Halliwell \cite{Hal99}). Note here that in the
standard (as opposed to the inverse to be introduced below)
formulation of the decoherent histories there exist more than one
(incompatible) decoherent set of histories, in general, and for
each of these sets, a set of records exists (incompatible with
those corresponding to other decoherent sets). The persistence of
these records in time is something not in general guaranteed.

To sum things up, in the standard histories approach
\begin{itemize}
\item Both the system and its environment are given. The latter is
represented by prescribing initial conditions and in some cases
final conditions.

 \item The space, in particular, its topological structure
is presupposed.

 \item The interactions are
given in terms of the decoherence functional, which encodes the
dynamical information. For the complete dynamics, the full
Hamiltonian must be known.
\end{itemize}

 In the \textbf{inverse histories} approach (or else `tomographic') developed in our previous
 paper \cite{case I} things are different. We solve the
\emph{inverse} problem. While in standard histories we are given

\begin{itemize}
\item the Hamiltonian

 \item initial conditions

 \item the space on
which they are defined
\end{itemize}

\noindent and the aim is to predict probabilities for histories,
we do the opposite thing. We have the relative frequencies
corresponding to different histories (belonging to a particular
decoherent set) and we consider the records in the `final' time
that are related to this histories\footnote{The existence of these
records is guaranteed by the relation of decoherence with records
as it is mentioned earlier.}. The decoherent set the histories
that we measure, belong, is determined by the `basis' we measure
the records in the final time. \ie we perform an actual physical
measurement in the Copenhagen sense, in the final time to get one
particular record (corresponding to the whole `generalized
trajectory' of our system). The basis we choose to measure the
`record space' \footnote{that physically could correspond to an
environment that produces decoherence for the system in question}
will single out a unique `preferred' set of decoherent histories.

To get the relative frequency of these histories (corresponding to
the probability of finding the one record or the other given a
fixed basis), we repeat the experiment. Then, by making certain
assumptions about these records,  we recover the topological
structure of the underlying spacetime. The assumptions are that
each record consists of sub-records which we may identify as
records of {\em events}.

Here we should stress that due to the above assumption that the
records capture the spatiotemporal properties of the system and
are thus records of events, the extended configuration space of
the system is identified as the \emph{`effective spacetime'}. This
identification is valid only under the above assumption about the
nature of the records considered.

What is done, is that from a set of events, with no other
structure presupposed (:{\it a priori} imposed from outside the
system), we end up with a causal set representing the discretized
version of the \emph{effective spacetime} the system in question
lies. We will then proceed to consider the topology of this
spacetime. We should stress here that the topology that we are
speaking here and that we will use in the rest of the paper,
concerns spatial topology, as it is usually understood. So in the
case that we will consider a `spacelike' surface, topology is
understood as the spatial one (and observables of this will be
things like the homology) while when we speak of `4-dimensional'
topology we will mean a series of `3-dimensional' topologies that
are ordered according to time (or more precisely a
parameter-time). In the latter case we can also speak of
transitions of one `3-topology' to another.

The spacetime that we get will be an `\emph{effective}' one, and
in a sense it accounts for certain properties of the Hamiltonian,
such as interactions with other objects not controlled by the
experimenter. For instance, the latter could be some kind of
`repulsive' field that prohibits the system to go somewhere (:in a
region of its configuration space), which can then be recovered as
a hole (:a dynamically inaccessible region) in that space. Note
that since no force propagate faster than the speed of light, the
causal structure will not be affected by any interactions and the
set of events would give us the effective spacetime.

Therefore in our set up we may carry out our experiment
sufficiently many times and we have access to the following two
things: (i) the set of possible histories and (ii) the relative
frequencies for each history to occur for every initial state.
From this we recover the parameters of the experiment.

Furthermore, we assume that \emph{the records capture the
spatio-temporal properties} (of the system in focus). This means
that \emph{the histories are coarse-grained trajectories} of the
system, belonging to a space whose topological properties we
ultimately wish to deduce. We shall then claim that the whole
concept of spacetime, as a background structure, does not make
sense in finer-grained situations. In this way, all the histories
are single-valued on our discretized version of `effective
spacetime'. One should note here that we may still have histories
that have the particle in a superposition of different position
eigenstates, but only if the latter are `finer' than the degree of
our coarse-graining. With the coarse-graining we effectively
identify (\ie we group into an `equivalence class' of some sort)
the points that we cannot distinguish operationally, with the
resulting equivalence class of `\emph{operationally
indistinguishable points}' corresponding to a `blown up', `fat
point' in our discretized version of `effective spacetime'.

\paragraph{Records and Sub-records.}
Here we should make a clarification about what the records are.
Each of these, corresponds to a total (generalized) trajectory of
the system and consists of many sub-records that correspond to
each of the points/events that constitute this trajectory. Having
got just a `frozen picture' (at the final time) of the trajectory
we are not able to say which event (sub-record) occurred before
and which after if we do not presuppose anything about the
dynamics and the background as was mentioned earlier.

In the rest paper when we will refer to records, we will mean
sub-records while for the full record we will refer to as a
history, causal chain or trajectory.

\paragraph{Effective Topology.} In our approach, we
consider the topology of the effective spacetime, which we derive
from our observations. Thus we may or may not assume the existence
of the spacetime with a certain topology. In either case we cannot
determine this \emph{`real'} topology from our measurements and we
therefore merely speak of \emph{effective topology}---the topology
of a model of spacetime which accords with our experiments and
fits their outcomes.

Consider an example. Suppose we have derived a non-trivial
topology for the `background' spacetime---say, for instance, that
it has a defect, such as a hole. This only indicates us that we
have non-contractible loops, nothing more. Why these loops fail to
be contractible---due to the existence of a `real hole', or
because of, say, the presence of a potential barrier---such a
question is, as a matter of principle, not verifiable within our
approach.

\paragraph{Operationalistic Setup.} Our approach is essentially
\emph{operationalistic}.  The set of records, is regarded as the
only source of information we possess about the system we wish to
explore. The effective topology then refers to the  effective
spacetime of the system in question. In our tomographic approach,
we are given the sets of observed histories together with their
relative frequencies, from which then we reconstruct the
parameters of the problem.

We assume that some of the records may be identified with
particular events, \ie spacetime `points'. Furthermore, we claim
that this is the only case we may speak of a background spacetime
proper. That is, if we do \emph{not} have access to events even in
principle, we \emph{cannot} speak about their support or their
topological and causal nexus, as, say, in the causal set scenario
(causet). Then, relative frequencies are recovered by repetition
of the whole histories involved: by restarting the system in an
identical environment and letting it evolve for the same amount of
time.\footnote{From our vantage, `\emph{history could in principle
repeat itself}' (pun intended).} In our operationalistic
(ultimately, relational-algebraic) view, the only way one can talk
about some background structure such as `spacetime', is relative
to something else. More precisely, we use our data (records) to
(re)construct an `arena' for a particular subsystem of the
universe that we are interested in, and it is \emph{only} in this
sense that we may speak of `spacetime'.

More precisely, we have \begin{itemize}
 \item[(a)] A system (call it `particle'),
which is placed into an appropriate experimental environment, and
we are able to repeat the experiment with \emph{the same} initial
conditions. In this way we get the relative frequencies of the
records.

We may also vary the initial conditions of the system in question,
leaving all the environment (and records) the same. For each
initial condition of the system, we rerun the experiment. These
first two steps give us the set of all possible histories
(coarse-grained trajectories) of the particle, as well as their
relative frequencies.

\item[(b)] The space of records. It is a space of data resulting
from controlled environment tampering with the system, and it is
supposed to capture its spatiotemporal properties. Records are
interpreted as \emph{distinguishable} spatiotemporally, events.
That is to say, that despite the fact that we do not know the
structure of the set of records that corresponds to events, we can
identify each record corresponding to a spacetime point as being
different from the others. Thus, while we know nothing {\it a
priori} about their causal or spatial (topological) ordering,
events can be labelled so that we do not have identification
problems.

 We can vary each record corresponding to a particular event
independently. The variation is in some sense small---this may be
effectuated by a `small energy' variation of the record. The
record can be thought of as a generalization of a `measuring'
device, and varying the record can be understood as varying this
device. This generalization is in the sense that the notion of
reords in the decoherent histories is believed to replace and
generalize the notion of measurement as was stated for example, by
Gell-Mann and Hartle in \cite{GH92} and by Halliwell in
\cite{Hal99}. The latter is assumed to be small enough not to
affect the `topology' of the records (\ie neighborhoods in the set
of records remain the same). By `topology' we mean a reticular
structure associated with appropriate coarse-graining of a region
of the effective spacetime we explore. The aforementioned
variations give us the proximity relations between events in the
classical case \cite{case I} and for the `statistical' recovery of
topology which will be described later.

\end{itemize}

Experiments are carried out repeatedly and multiply. We label the
runs by initial conditions of the system, number of run and
`positions' of events.\footnote{By this we mean whether or not we
varied one record corresponding to an event.} Each run gives us a
history, \ie a sequence of causally related events that in the
relativistic case will correspond to a causal chain.

The records correspond to different repetitions of the same
experiment (of non-trivial temporal `width'), but could also
correspond to simultaneous measurements of different systems (that
have evolved) and where initially in the same state. One could
imagine such a scenario considering the data from different angles
of the CMB (Cosmic Microwave Background) presupposing isotropic
universe.

To conclude, from our experiments we get the following
information:

\begin{enumerate}

\item The set of histories of the system associated with a fixed
set of initial conditions. We call this set of histories {\dff
fiducial set}. Here we emphasize that these correspond to
coarse-grained `trajectories'\footnote{The inverted commas are
added to the word `trajectories', since the space on which they
are defined is not presupposed.}. We denote the \textbf{set of all
histories} to be $\histories$, while each history that is
contained in it is denoted by $\history_i$. Note that within these
histories-`trajectories' the order of the events is not known. The
set of all possible events, or else the set of `spacetime' points
will be denoted by $\causet$.

\item The relative frequencies of outcome of these histories
depending on the initial conditions. This is a function $$f_j:
\histories\rightarrow[0,1]$$ which gives the relative frequency of
histories for each particular initial condition (corresponding to
$j^{th}$ initial state of the system).

\item The change in the relative frequencies when one event is
varied. This is a function  $$f^p_j: \histories\rightarrow[0,1]$$
which is the \emph{new} relative frequencies when the event $p$
has been varied. This will lead us to the statistical way to get
proximity relation between the points produced by the fiducial set
of histories.

\end{enumerate}

It is important to note that we already have the fiducial set of
histories \emph{before} we vary the records. This fiducial set of
histories provides us the set on which the topology is imposed.

In \cite{case I} we used all that to get as much information about
the underlining topological space as possible in the non
relativistic case. More precisely, we were able to recover the
number of components the effective spacetime had and the number of
components a spatial surface had from purely algebraic
considerations, not using the third of the above mentioned
information. We were also able to recover the topology of the
 background effective spacetime, using the proximity relation on spatial
surfaces derived from the change of relative frequencies. The
latter was named \textbf{statistical approach}.

\section{`Relativistic' case}\label{srelcase}

When we dwell on the relativistic case, the speed of the light is
the upper bound $c$ in the transmission of (material) information,
this further restricts the set of possible histories-trajectories.
The restriction is simply that they need to be causally related,
which gives rise to a partial order in the set $\causet$.

Here we should point out that our construction lies in the
`organization' of some records that are physically measured. This
implies, that in some sense we do have a preferred frame, namely
the one that the observer (us) that measures the records, lies.
But what we wish to highlight is something different: once the
records have been collected, we reconstruct an effective causal
space (:`time-space') structure, and only after this is done we
can study its (again effective) transformation theory
(:relativity). The latter is already built into the `kinematical'
variation of records technology that we suggest.

Now we will first consider the case where we are given a partially
ordered set as our effective spacetime. Later we will come back to
the case where the relation between the elements of the set
$\causet$ is unknown, but we do know the set of possible histories
(causal curves) in the form of a covering of the set $\causet$
with subsets $\chn^i$. The set of all these is $\chns=\{\chn^i\}$.
In the sequel, we will restrict our attention to histories that
are causal curves. Our considerations are similar to those that
led us to consider trajectories in the classical case (see also
comment in introduction before the records and sub-records
paragraph). We should stress here, once more, that in the context
of our set up the histories-chains are a covering with subsets of
$\causet$ that within each of these subsets the order of the
events is unknown.

\paragraph{Partially ordered sets.} A partially ordered set,
usually abbreviated as {\dff poset}, is a set $\causet$ endowed
with a relation $\preceq$ having the following properties:

\begin{itemize}

\item Reflexivity: $\forall p\in\causet\quad p\preceq p$.

\item Transitivity: $\forall p,q,r\in\causet\quad \,p\preceq q,\;
q\preceq r \Rightarrow p\preceq r$.

\item Antisymmetry: $\forall p,q\in\causet\quad \,p\preceq q,\;
  q\preceq p \Rightarrow p=q$.

\end{itemize}

A subset $\chn\subset\causet$ is called a {\dff chain} (also known
as a {\dff linearly} ordered subset) if any pair of its points is
ordered: $\forall p,q\in\chn\quad p\preceq q$ or $q\preceq p$. In
the sequel, we shall consider maximal (that is, inextensible)
chains in $\causet$ and we will denote the set of all maximal
chains by $\chns$:

\begin{equation}\label{edefmaxchains}
  \chns
  \;=\;
  \{\mbox{ maximal chains of $\causet$ }\}
\end{equation}

In a similar way, we define an {\dff antichain} to be a subset
$\antich$ of $\causet$ such that no pair of its points is ordered:
$\not\!\exists p,q\in\antich\quad p\preceq q$. We shall need
maximal antichains in $\causet$, and denote the appropriate set by
$\antichs$:

\begin{equation}\label{edefmaxantichains}
  \antichs
  \;=\;
  \{\mbox{ maximal antichains of $\causet$ }\}
\end{equation}

\subsection{Causets}\label{scausets}

Discretized spacetimes with an `inherent' causal structure are
referred to as {\dff causal sets}, or causets for short. Causets
are partially ordered, locally finite sets. The points of causets
are thought of as spacetime points (:events). Local finiteness
represents the (supposed!) fundamentally discrete nature of
spacetime. It has been developed as a possible alternative to the
spacetime continuum (:manifold) of General Relativity
\cite{Sor91a,Sor91b,prg}. Like in General Relativity, in every
causet we can define both future $J^+(p)$ and past $J^-(p)$ cones
for each of its events ($p\in\causet$):

\begin{equation}\label{edeffutpas}
  \begin{array}{l}
    J^+(p)\;=\;\{q\in\causet\mid p\preceq q\}
    \\
    \\
    J^-(p)\;=\;\{r\in\causet\mid r\preceq p\}
  \end{array}
\end{equation}

An element of a $\causet$ is said to be minimal if $J^-(p)=\{p\}$,
and maximal when $J^+(p)=\{p\}$. The notions of future and past
cones can be extended to subsets of $\causet$. For
$A\subseteq\causet$

\[
  \begin{array}{l}
    J^+(A)\;=\;\{q\in\causet\mid \exists a\in A\quad a\preceq q\}
    \\
    \\
    J^-(A)\;=\;\{r\in\causet\mid \exists a\in A\quad r\preceq a\}
  \end{array}
\]

In terms of posets, the chains stand for causal curves, while the
antichains are reticular analogues spatial (hyper)surfaces. A
foliation $\foliation$ is a partition of a causet $\causet$ into
spatial surfaces (\ie antichains) which respects the partial order
$\preceq$ in $\causet$, namely,

\[
\forall A,B\in\foliation \quad A\cap J^+(B) \neq\emptyset
\;\Rightarrow\; A\subseteq J^+(B)
\]

Starting from this, we may introduce an ordering $\sqsubseteq$ on
$\foliation$; namely, for $A,B\in\foliation$

\begin{equation}\label{edeforderfol}
  A\sqsubseteq B
  \;\Leftrightarrow\;
  A\subseteq J^+(B)
\end{equation}

A discrete analogue of a globally hyperbolic spacetime is a
linearly foliable causet, \ie when the order $\sqsubseteq$ is
linear (see definition above). It can be shown that any past
finite causet (and this is the case we presently consider) admits
a linear foliation, while this is not the case for a spacetime
which admits closed timelike curves. The following construction
proves this statement:

\begin{itemize}
  \item $A_0:=\{\mbox{minimal elements of $\causet$}\}$
  \item $A_1:=\{\mbox{minimal elements of $\causet\setminus A_0$}\}$
  \item $A_k:=\left\{\mbox{minimal elements of $\causet\setminus
    \left(\bigcup_{j=0}^{k-1}\limits{}A_j\right)$}\right\}$
\end{itemize}

Here we should note the existence of one `preferred' foliation in
each past-finite causet, which is the one derived from the above
mentioned procedure. In this foliation, each event is in the
$n^{th}$ surface\footnote{The antichains obtained in this way are
not necessarily maximal in $\causet$. For instance, in the example
presented in Section \ref{Ambiguities in Causet} we have
$A_4=\{6\}$, which is not maximal in $\causet$ as it can be
augmented to, say, $\{5,6\}$.} where $n$ is the maximum number of
steps to reach to the event by following a causal curve. So, the
r\^ole of spacelike surfaces in our approach is played by
antichains in $\causet$ that belong to $\foliation$.

\subsection{Reconstruction of causal sets}\label{semergcaus}

As described above, our experiments provide us with causets that
are only unstructured collections of points $\events$ (:events),
without telling us anything about their partial order. However, we
know histories which are maximal causal chains (denote them
$\chns=\{\chn^i\}$).

For a time being, recall the classical case, where all histories
that involved one event in every spatial surface, were possible.
So we could have a history that in two consecutive instants have
events infinitely `far' spatially. Thus the information that two
points belonging to two consecutive spatial surfaces could be in
the same history added no information about their `spatial'
proximity and we were therefore unable to deduce more things about
the topology other than the number of components. We were forced
to use further measurements and recover the proximity with the
statistical approach. In the relativistic case, in contrast, the
upper limit in the transmission of information can provide us with
extra information and we will be able to recover more things
merely from the fiducial set.

Returning now to the relativistic case, we have a set and its
covering by subsets. The reconstruction procedure looks like the
following branching process \cite{rbr}.

\begin{itemize}

\item[Step 1. ] Pick a \textbf{maximal} collection of points
$\antich^0\in\events$ such that no pair of points
$p,q\in\antich^0$ belong to a chain $\chn^i$. This $\antich^0$
will be the set of minimal elements. Assign $i:=1$.

\item[Step 2. ] Consider the set
$\events^i=\events\setminus\antich^{i-1}$.

\item[Step 3. ] Pick a \textbf{maximal} collection of points
$\antich^i\in\events^i$ such that no pair of points
$p,q\in\antich^i$ belongs to a chain $\chn^i$ (of the initial
event set $\events$!). This $\antich^i$ will be the second layer,
if it exists, and is assigned $i:=i+1$. Then go to Step 2. If such
$\antich^i$ does not exist, the branch fails and one should
restart from Step 1.

\item[Step 4. ] First check for non-appearance of extra chains. If
it turns out that the foliation involved gives rise to a new
chain, the branch is rejected. Then return to Step 1, restarting
with a different maximal antichain. To see an example of the
appearance of an `extra chain', see below.

\item[Step 5. ] If the set $\events$ is exhausted and all the
causal chains can be reproduced without emergence of a `new' one
(see an example below), then the collection
$\antichs=\{\antich^i\}$ forms the foliation of $\events$, the
latter regarded as a causet proper\footnote{The end-product of
this algorithm is guaranteed to be a foliation since all posets
involved are foliable (see section \ref{scausets}).}.
\end{itemize}

What we have effectively done in the foregoing is the following.
We picked a partition of $\causet$ into anti-chains. To define the
anti-chains we used the set of causal chains---histories. We then
chose randomly an order on these anti-chains. After that, we
checked that our construction did not produce any new histories
(extra chains, in other words). If it did, we restarted the
procedure.

\paragraph{Chains and cones.} The set $\chns(p)$ is the union of
all maximal chains containing a point $p\in\causet$. For an
arbitrary point, we have $\chns(p)=J^+(p)\cup J^-(p)$; thence, for
minimal elements (for which $J^-(p)=\{p\}$)

\[\chns(p)=J^+(p)\]

The way to derive histories is the following. We pick a point in
$S^0$ and see which points are causally connected with it in
$S^1$. Then, we continue with the point we chose from $S^1$ and do
the same with $S^2$. Note that if we had chosen the correct
foliation we would not have obtained new chains, because of
causality's transitivity.

\paragraph{An example of `new' chain.} Here we show how `extra
chains', not existing in the initial poset, may emerge during the
reconstruction procedure described above in step 4. Consider the
poset $\causet$

\par

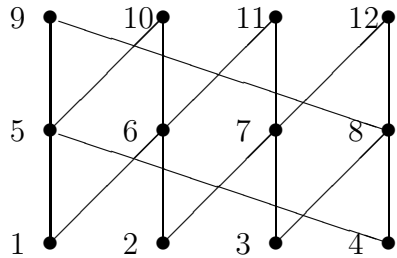
\unitlength1.5mm
\newcounter{ptlabels}

\begin{center}
\begin{picture}(30,20)
\multiput(0,0)(10,0){4}{\circle*{1.2}}
\multiput(0,10)(10,0){4}{\circle*{1.2}}
\multiput(0,20)(10,0){4}{\circle*{1.2}}
\multiput(0,0.5)(10,0){4}{\line(0,1){9}}
\multiput(0.35,0.35)(10,0){3}{\line(1,1){9.5}}
\multiput(0,10.5)(10,0){4}{\line(0,1){9}}
\multiput(0.35,10.35)(10,0){3}{\line(1,1){9.5}}
\multiput(0.75,9.65)(0,10){2}{\line(3,-1){29.5}}
\multiput(-4.5,-1)(10,0){4}{\stepcounter{ptlabels} \theptlabels}
\multiput(-4.5,9)(10,0){4}{\stepcounter{ptlabels} \theptlabels}
\multiput(-4.5,19)(10,0){4}{\stepcounter{ptlabels} \theptlabels}
\end{picture}
\end{center}

\medskip

\noindent and try to restore the order starting from $\{5,6,7,8\}$
as the set of minimal elements. So, if $5$ is a minimal element,
then $J^+(5)=\chns(5)=\{1,4,5,9,10\}$. Then take the element $1$
(for which we deduced $5<1$). In the set
$\causet\setminus\{5,6,7,8\}$, consider $J^+(1)=\{1,9,10,11\}$,
hence $1<11$. Thus, the chain $\{5,1,11\}$ must exist, but
actually it does not(!), therefore we reject the initial
supposition that the antichain $\{5,6,7,8\}$ is minimal.

\section{Ambiguities in Algebraic Causet
Construction}\label{Ambiguities in Causet}

 Note that in the above
way, we will eventually recover a foliation with some ambiguities,
\ie we will not get a unique partial order. The different causets
we will get will  be related to each other by some `symmetry'
transformations. One obvious would be an overall flip in
direction. Another one would be related with points that exist in
all histories and could be thought of as one-point spacelike
surfaces. The order this `surface' would exist, is ambiguous. To
further classify these ambiguities, let us proceed first to some
definitions.

\paragraph{Definition 1a:} Let A be a subset of $\causet$.
We define the `initial surface' of A, $S_i^A$ to be  the set of
minimal points of the subset A when considered as a partially
ordered set with respect to the partial order induced from
$\causet$.

\paragraph{Definition 1b:} We also define the `final surface' of A, $S_f^A$   to be the
set of maximal points of the subset A when considered as a
partially ordered set with respect to the partial order induced
from $\causet$.

\paragraph{Definition 2:} We define a point $q$ in a partial
order $\causet$ to cover $p\in\causet$ to mean that $q$ is in the
future of $p$  ($q\succeq p$) and $\nexists \quad r\in\causet
\quad\mid\quad q\succeq r\succeq p$.

\paragraph{Definition 3:}Transitive closure of a point $p$
in a partial order $\causet$ is the set of points $q\in\causet$
such that there exists a sequence of chains $\{C_j\}$\footnote{j
going from $0$ to $n$} with $p\in C_0$ , $q\in C_n$ and $C_k\cap
C_{k+1}\neq\varnothing\quad\forall\quad k\in[0,n]$.

\paragraph{Definition 4:}One-component subset, is a subset $A$
of $\causet$, that when consider as a partially order set from the
induced from $\causet$ order, it has only one component, \ie the
transitive closure in A, (when considered as partial ordered set)
of any point $p\in A$ is the set A itself.

One consequence of the above is that in a one-component subset A,
every subset of the final surface, $D\subseteq S_f^A$, has in its
past at least a point $p$ that has in its future points in the
complement of $D$ in $S_f^A$, $D^c$. A similar statement holds for
the initial surface $S_i^A$.

\paragraph{Definition 5:} We define a  subset A of
$\causet$ to be `complete' if \footnote{The following is
equivalent with the condition: $\forall\quad p,q\in A,\quad
J^+(p)\cap J^-(q)\subseteq A$ and $J^-(p)\cap J^+(q)\subseteq A$.}
$A=J^-(S_f^A)\cap J^+(S_i^A)$

\paragraph{Definition 6:} We will call the subset $A$
\emph{`information closed'} (abbreviated as \emph{i.cl.}), if it
has the following properties:
\begin{itemize}

\item[(a)] $\forall\quad C_j\in \mathcal{C}\mid\exists \quad p\in
S_i^A$ and $p\in C_j \Longrightarrow \exists\quad r\in S_f^A$ and
$r\in C_j$

\item[(b)] $\forall\quad C_j\in \mathcal{C}\mid\exists \quad p\in
S_i^A$ and $p\in C_j \Longleftarrow \exists\quad r\in S_f^A$ and
$r\in C_j$

The above conditions means that any `ray' (causal chain) that
enters the initial surface will cross the final AND any `ray' that
crosses the final has also crossed the initial. This would mean
that no `information' from other places of $\causet$ enters or
leave.

\noindent Note also that when the subset considered is a
`complete' one, then the following condition is equivalent with
(a) and (b):

$\left(J^-(S_f^A)\setminus J^-(S_i^A)\right)\cup
S_i^A=\left(J^+(S_i^A)\setminus J^+(S_f^A)\right)\cup S_f^A=A$

\noindent Otherwise this condition gives the completion of the
subset A (\ie the smallest complete subset that contains A).

\noindent Finally any $p$ in a complete \emph{i.cl.} subset that
does not belong to the initial or final surface is covered only by
points in the subset \footnote{That is true, since  there is no
`ray' escaping the final surface (i.cl.) and there are no `holes'
since the subset is complete.}.

\end{itemize}

\paragraph{Definition 7:} We define a `flip' of a subset A of a
partial order $\causet$ to be a new partial order $\causet'$
having the same elements as before, and the relations between the
points to be as follows:
\begin{itemize}
\item[(a)] All the points  in $\causet'\setminus A$ have between
them the same relation as in $\causet\setminus A$ , \item[(b)] The
relation of any point $p\in\causet '\setminus A$ with any point
$q\in A$ is the same as the relation of the same point $p$ now
belonging to $\causet\setminus A$ with the point $q\in A$.
\item[(c)] The relation of points $p,q\in A$ when seen as subset
of $\causet'$ the new partial order, is the opposite of the
relation between the same points $p,q$ when seen as point
belonging to a subset of the old partial order $\causet$.
\end{itemize}

This condition already puts certain constrains, since the partial
order is transitive, and we do not want to alter the relation of
two points outside $A$ and we also should not make any closed
loop.

Note here that we could have considered a more `relaxed'
definition of `flipped' subsets that we did not require the
condition (b). In that case we would have allowed to flip the
relation of some points in the subset with some other points that
are outside imposing that the relation of the point outside the
subset with other points outside, does not change. This is exactly
as if we had included the point in the subset but required the
condition (b) for flipping a subset to hold. This suggests that we
can get all the possible `inversions' of subsets of the `weaker'
condition, by some `flips' of the kind defined in definition 7 and
therefore this definition is general enough.

\paragraph{Definition 8:} A subset $A$ of a partial order $\causet$,
is called \emph{`invertible'}, if we can `flip' the overall order
of the subset $A$ without altering the set of `causal chains' of
$\causet$\footnote{This condition adds more restrictions to the
allowed `flips'.}.

\begin{theo}\label{flips} A one-component subset $A$ of $\causet$
is `invertible' if  and only if it is:
\begin{itemize}
\item[(a)] `Complete'. \item[(b)] Information closed. \item[(c)]
If $p\in\causet$ covers $q\in S_f^A \Rightarrow $ $p$ covers
$r,\quad \forall \quad r\in S_f^A$. \item[(d)] If $q\in S_i^A$
covers a point $p\in\causet$ implies that all $r\in S_i^A$ covers
$p$.
\end{itemize}

\end{theo}

\begin{proof} We will first show that if a subset $A$ of $\causet$
obeys the conditions (a)-(d), implies that $A$ is `invertible', by
construction.

By the condition that the subset $A$ is complete and information
closed we know that the only `direct' links between points of $A$
and points of $\causet\setminus A$ are those of $S_f^A=\{l_i\}$
with the points $\{p_i\}$ that cover them and those of
$S_i^A=\{r_i\}$ with the points $\{q_i\}$ that are covered by the
$r_i$'s (see also final comment at definition 6). We could now
`cut' these direct links `invert' the order of relations in subset
$A$  and then join back $A$ in such a way that  all the points of
$S_i^A=\{r_i\}$ are covered by all the points $\{p_i\}$ and the
points of $S_f^A=\{l_i\}$ cover all the points $\{q_i\}$.
Condition (c) and (d) guarantee that this will not produce new
chains, since all `sub-histories' in A are linked with all the
$p_i$'s and $q_i$'s. In this way we will end up with a partial
order $\causet'$ that has the same set of chains with $\causet$
with the subset $A$ having the opposite relations in $\causet'$
and the points in the rest set having the same relations between
them and between them and points of $A$ as required by definition
7.

We now proceed to prove the converse,showing that each of the
conditions (a)-(d) are necessary conditions.

(a) If a subset $A$ is not `complete', this means that: $\exists
p\in\causet\setminus A\mid\quad r\succeq p\succeq q$ for $q,r\in
A$. Inverting the relation between $q$ and $r$ implies that
$q\succeq r$ which makes impossible for $p$ to be in the future of
$q$ and in the past of $r$ without creating a closed loop. So
according to the condition (b) of definition 7 the subset $A$
cannot be flipped.

(b) If a subset $A$ is not information closed this means that
there exists at least a `ray' passing from the subset and either
escaping the final or the initial surface. Let $p\in A$ be the
last point in $A$ of the escaping ray, and assume, for the moment,
that the ray escapes to the future and $q\in\causet\setminus A$
covers $p$. If the point $p$ belongs to the final surface $S_f^A$
we consider another ray, since this ones does not contradict the
definition of information closed (rays do escape from final
surface to the future and from the initial to the past). There
exists at least one chain containing $p,q$ and no other point in
the future of $p$ in $A$. By inverting the subset $A$, every
history containing $p,q$ will also have points from what used to
be the `future' of $p$ in $A$ and therefore the chain that we
mentioned before would not exist and the subset $A$ is not
`invertible'. Note that if the ray escapes to the past, similar
argument holds, considering the initial surface instead of the
final and the past of $p$ instead of the future.

(c) If the condition (c) did not hold, it would mean that there
exists a point $p\in\causet$ that covers a subset of the final
surface $D\subseteq S_f^A$ and does not cover the complement of
$D$ in $S_f^A$, $D^c$. This would mean that either there doesn't
exist a chain including $p,q_j$ where $q_j\in D^c$ or there
doesn't exist a chain with $p,q_j$ and no other element of
$\causet\setminus A$ in between $p$ and $q_j$. On the other hand,
there exists a chain with $p,r_i$ where $r_i\in D$ with no  other
element of $\causet\setminus A$ in between $p$ and $r_i$.

Inverting subset $A$, if we want to have a chain with $p,r_i$
where $r_i\in D$ with no  other element of $\causet\setminus A$ in
between $p$ and $r_i$, we will have to create at least one chain
including $p$ and $q_j$ with no other element in between. This is
due to the fact that subset $A$ being one-component subset, it
necessary has the property that in the past of points $r_i\in D$
there exists at least one point that has in its  future a point in
$D^c$ (see comment after definition 4). Thus inverting the
relations in the subset will bring to `same' fate the point in $D$
and the point in $D^c$ that have in their past the point
connecting them. We would have thus, created a new chain and the
subset $A$ would not be `invertible'. Therefore it has to obey
condition (c).

(d) This can be proven similarly to (c). Note that if we invert
all the relations in $\causet$ condition (d) becomes condition (c)
and since the set of histories is clearly not affected by an
overall flip this condition should also hold.

This completes the proof.
\end{proof}

A direct consequence of the above theorem, is that if a subset is
information closed and $(S_i^A\subseteq S_i^{\causet}$ or
$S_i^A=\{p\}$ for some $p\in\causet )$ AND $(S_f^A\subseteq
S_f^{\causet}$ or $S_f^A=\{q\}$ for some $q\in\causet )$ then the
subset is invertible. This is due to the fact that the previous
condition is just a special case of the theorem.

If we want to consider a subset that has more than one components,
we treat each component separately.

We furthermore speculate, that any ambiguity in the causet
construction of section \ref{semergcaus} is due to some ambiguity
of the direction of some subsets. This is natural to assume, since
the information about the direction is not given from the set of
histories-causal chains.

\begin{conj}\label{flips2} We can transform any partial order to another
with the same set of chains (when considered as subsets with no
order) by some combinations of flips of one-component subsets.
\end{conj}

Here we should also note that  a Minkowski space (where
information `spreads' in space from every point) or actually in
any space having that feature, there wouldn't be neither any
non-trivial \emph{i.cl.} subsets nor any subset obeying conditions
(c) and (d), and furthermore we suspect (if conjecture
\ref{flips2} is true) that we wouldn't have any ambiguity apart
from the overall flip or else `time-reversal'\footnote{Note that
the general features described above are not necessarily satisfied
by our operationalistic `effective' spacetime. }.

Let us now explore an example below to demonstrate the above.

\paragraph{An example.} Consider the poset $\causet$:
\par
\unitlength1.5mm
\newcounter{plabels}

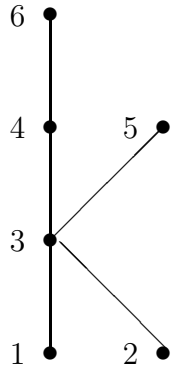
\begin{center}
\begin{picture}(30,30)

\multiput(0,0)(10,0){2}{\circle*{1.2}}
\multiput(0,10)(10,0){1}{\circle*{1.2}}
\multiput(0,20)(10,0){2}{\circle*{1.2}}
\multiput(0,30)(10,0){1}{\circle*{1.2}}
\multiput(0,0.5)(10,0){1}{\line(0,1){9}}

\multiput(0,10.5)(10,0){1}{\line(0,1){9}}
\multiput(0,20.5)(10,0){1}{\line(0,1){9}}
\multiput(10.35,0.35)(10,0){1}{\line(-1,1){9.5}}
\multiput(0.35,10.35)(10,0){1}{\line(1,1){9.5}}
\multiput(-4.5,-1)(10,0){2}{\stepcounter{plabels} \theplabels}
\multiput(-4.5,9)(10,0){1}{\stepcounter{plabels} \theplabels}
\multiput(-4.5,19)(10,0){2}{\stepcounter{plabels} \theplabels}
\multiput(-4.5,29)(10,0){1}{\stepcounter{plabels} \theplabels}
\end{picture}
\end{center}

\medskip

Assume we just have the set of chains-histories. Try to restore
the partial order, starting with $\{3\}$ as the set of minimal
elements $S^0$. Set as second layer the surface $S^1=\{4,5\}$, as
third $S^2=\{6\}$, and last, the set $S^3=\{1,2\}$. Now
$J^+(3)=\{1,2,4,5,6\}$. Choose one of the points in the future of
3 that are in $S^1$---say for example, choose $4$ so that we have
$3\preceq 4$. Pick one point in $S^2$ that is causally connected
to $4$. This is the point $6$. Repeat this for the last layer and
end up with the two histories $\{3,4,6,1\}$ and $\{3,4,6,2\}$,
that exist. Now choose the element $5$ from $S^1$. Then, there
does not exist any point in $S^2$ that belongs to the same history
with $5$ (since $6$ is the only element there, and it is not
connected to $5$). We continue with the next surface $S^3$. With
the procedure described we have recovered the following histories:
$\{3,4,6,1\}$, $\{3,4,6,2\}$,$\{3,5,2\}$, and $\{3,5,1\}$. We have
thus recovered all the histories, no matter that it is \emph{not}
our initial causet.

\paragraph{`Wrong'} poset $\causet'$:
\unitlength1.5mm
\newcounter{plabels2}

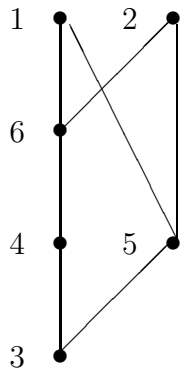
\begin{center}
\begin{picture}(30,30)

\put(0,0){\circle*{1.2}} \put(0,10){\circle*{1.2}}
\put(10,10){\circle*{1.2}} \put(0,20){\circle*{1.2}}
\put(0,30){\circle*{1.2}} \put(10,30){\circle*{1.2}}

\put(0,0.5){\line(0,1){9}} \put(0,10.5){\line(0,1){9}}
\put(10.35,10.35){\line(0,1){19}} \put(0,20.5){\line(0,1){9}}
\put(0,0.35){\line(1,1){9.5}} \put(10.35,10.35){\line(-1,2){9.5}}
\put(0.35,20.35){\line(1,1){9.5}}

\put(-4.5,-1){$3$} \put(-4.5,9){$4$} \put(-4.5,19){$6$}
\put(-4.5,29){$1$} \put(5.5,9){$5$} \put(5.5,29){$2$}
\end{picture}
\end{center}

\medskip

The important point of this ambiguity is that the surface
$\{4,5\}$ is before the $\{1,2\}$ while normally should be the
other way round.

From our initial causet we could end up to the one just described,
by some transformation related with the fliping of some subsets
that are \emph{`invertible'}. First we  have an overall flip to
end up with $S^0=\{5,6\}$ , $S^1=\{4\}$ , $S^2=\{3\}$ and
$S^3=\{1,2\}$. Then we consider the subset $A=\{3,4,5,6\}$ that is
`complete', \emph{i.cl.} and obeys conditions (c) and (d). This is
true, since 6 and 5  have no point covering them in the past, and
point 3 being one point surface and also the final surface of
subset A is covered by 1 and 2 and obeys (c). This , by theorem
\ref{flips}, means that it is \emph{`invertible'}. We flip the
subset A and end up with $S^0=\{3\}$ , $S^1=\{4,5\}$ , $S^2=\{6\}$
and $S^3=\{1,2\}$ which is the false causet that we got previously
being consistent with our fiducial set of histories.

We  should note here, that if instead of the subset A we took
subset $B=\{3,4,5\}$ it would still be \emph{i.cl.} but  the
condition (d) would NOT be satisfied ($S_i^A=\{4,5\}$ and point  6
is covered by point 4 but not by point 5. The subset would still
be in some sense invertible (not according to definition 7, but
with the weakened definition 7 dropping condition (b)). We could
put 6 in the future of the subset and join it with point 4, while
we could keep the links of 6 as they were (1 and 2 covering 6) and
then link 5 to 1 and 2. This would have altered the relation of
point 6 that is not in the subset under consideration B, with
points in the subset. This would give us $S^0=\{3\}$,
$S^1=\{4,5\}$ , $S^2=\{6\}$ and $S^3=\{1,2\}$ which is the `false'
causet we got. The important point here is that there isn't a need
to weaken the definition of `flips' to allow such inversions,
since we could get this by just containing the point 6 in the
subset, to get subset $A$ that gives the same result (see also
comment after definition 7).

The bottom-line is that ambiguities that cannot be resolved by
knowing all the histories are `intrinsic', and there is no
physical argument for us to believe that we are in the one or the
other causet. We could either say that we are in a superposition
of causets (under the proviso that it is possible with further
`measurements' to determine which one we are in), or that these
causets are operationally equivalent (:indistinguishable).

In our operationalistic approach, we may claim that if from our
measurements we cannot distinguish sharply a causet that is in
force, then our system is most probably described by a
\emph{superposition} of causets. This is in accordance with the
idea that when a measurement is made, the state `reduces' to the
projection on the total subspace that we measured, rather than to
the projection to a particular one-dimensional subspace . The
other possibility is to regard the `physical' states as being
\emph{equivalence classes} of causets with the equivalence
relation being the existence of the same set of histories
realizing them. In the later case though, given the fact that we
expect the topology of those causets to be non-homeomorphic, it
would be difficult to make any meaningful statement about the
topology of the corresponding effective spacetime (see also
section \ref{stransprob}).

We have showed how to reconstruct the partial order of the
effective spacetime in the relativistic case, in a combinatorial
way and abide by (\ie they can be checked according to) the
following {\dff consistency principle}:

\begin{quote}

Choose a partial order $\preceq$  on the set $\events$ of events.
Then we make a working hypothesis: `\emph{the partial order
$\preceq$ does not contradict our experiments}'. To accept or
reject this hypothesis, we just build the set of all maximal
chains $\chns(\events,\preceq)$. Then, if
$\chns(\events,\preceq)=\chns$, we accept the hypothesis;
otherwise we reject it.\footnote{Generically, the causal order is
reconstructed up to the ambiguities described above. The
corresponding topologies are in general different
(non-homeomorphic).} Note that the existence of a new chain,
immediately contradicts this consistency criterion.
\end{quote}

\noindent At this stage we have the causet associated with our
laboratory.

\section{Recovering the topology: statistical vs algebraic
approach}\label{srecspatop}

So far we have only used the set of all histories, while the
relative frequencies have not yet been used. We shall now consider
ways to recover some topology on this causet. Here we should
remind the reader that when we speak of topology, we mean the
`spatial' topology in the way that is usually understood.
Observables of this would be things like the homology or other
topological invariants. When we will speak of the topology of the
spacetime, we will mean topology of `3-dimensional' spatial
surfaces patched together according to an ordering. This ordering
is according to the, (unphysical) parameter-time.

There are two ways to recover the topology. The first one is to
vary the records as it was done earlier, in the classical case
\cite{case I} and we will call it the \textbf{statistical} way of
recovering topology while the second uses merely the derived
causet as its only source of information and will be referred to
as \textbf{algebraic} way of recovering topology.
\subsection{Statistical approach}
 We have the relative frequencies of each
history $C_i$ with initial condition `$j$', labelled $f_j(C_i)$,
and the relative frequencies having varied point `$p$', labelled
$f^p_j(C_i) $ (see final part in Introduction) . We then take a
small positive number $\epsilon\ll 1$. We define another function,
the difference function, as follows:

\begin{equation}
\delta^p_j: \histories \rightarrow [0,1]: \mid
f_j(C_i)-f^p_j(C_i)\mid
\end{equation}

\noindent We then consider all the points belonging to the
histories $C_i\in\mathcal{C}$ that

$\delta^p_j(C_i)>\epsilon$. We name them j-neighbors of $p$.
Physically we assumed that the relative frequencies of histories
containing points close to the one we vary, will alter more than
histories containing points only far from the point in question
(spatially and temporally). So we have:

\begin{equation}
q\in N^p_j\Longrightarrow \exists \quad q\in
C_i,C_i\in\histories\mid \delta^p_j(C_i)>\epsilon\nonumber
\end{equation}

We then consider different initial conditions `$j$' and we group
all the neighbors together to form the neighbors of `$p$' , $N^p$.
\begin{equation}
q\in N^p\Longrightarrow \exists\quad j\mid q\in N^p_j\nonumber
\end{equation}

We define spatial neighbors of the point `$p$' those points that
are in  $N^p$ but do not belong to the any history containing
`$p$'.

\begin{equation}
SN^p\mid q\in[N^p\setminus\cup_i C_i]\quad, \quad p\in
C_i\quad\forall\; i\nonumber
\end{equation}

By repeating this for every point in each of one spacelike surface
we may recover the proximity and therefore the topology of this
slice in the usual way-\eg as it is done in metric spaces.

 We will have obtained the topology of each
spatial components. We can then choose an arbitrary partitioning
of these slices to get the total `4-dimensional case' where we
will be able to see transitions from different
topologies\footnote{We consider `effective spacetime' and thus, we
are expected to see topology changes.}. We then check that we do
not have contradiction.This contradiction could be due to, for
example, some event being affected by a change in an event to its
future rather than to its past (:`advanced' and `retarded'
contradiction, respectively).If a contradiction arises, we pick
another `partitioning', so on and so forth, until the correct one
is obtained. In this way, previous ambiguities in the causet
construction, such as those related with an overall flip would be
resolved. So in the previous example in section \ref{Ambiguities
in Causet} the ambiguity would be resolved, since it contained an
overall flip. In other ,limited, cases, we would still have
ambiguities . In particular the order of two points $p,q$ that
belong to an `invertible' subset A of $\causet$ that consists of a
single chain (\ie a chain that is \emph{i.cl.} and `complete')
would still be ambiguous if $p,q\notin S_f^A$. If they were in the
final surface we could see that `varying' them affected the
relative frequencies of the next surface points (provided it is
not a single one). If none of them were in the final surface,
their order would remain ambiguous.

\subsection{Algebraic approach}\label{salgrectop}

We will assume for the moment that we have a unique unambiguous
causet. In this case there exist already certain way to speak
about topology (and certain geometrical properties furthermore).

Following \cite{Bri-Gre91} we may define some notion of distance
for timelike separated events, say $p\preceq q$ to be the number
of maximum steps in the partial order one needs to travel to go
from the one, $p$ to the other, $q$. This would correspond to
`proper time'. We then proceed to define distance of two points
being spacelike separated by considering the following. The only
way for an `inertial' observer in one point to know about its
distance to another that is spacelike separated is by considering
standard clocks and light beams. We would be therefore interested
in the distance of a point from a geodesic corresponding to a
history. We consider a point $x$ and a geodesic $C$ such that $w$
and $z$ are points of $C$ such that $w\preceq x\preceq z$. For
point $x$, let $l(x)$ be the highest point in $C$ which is below
$x$, and $u(x)$ the lowest point of $C$ that is above $x$. Then
$d_s(x,C)=d(l(x),u(x))/2$ where $d(. ,.)$ is the proper time.

Using those concepts one may define neighborhoods on the spatial
surface by choosing a particular distance around every point (c.f.
balls in usual metric spaces).

Alternatively one could follow \cite{Major:2005fy} and thicken
every anti-chain by considering the immediate future. Then use
this to define some `shadows' on the initial anti-chain, that
would group together points to form a finite cover of this set, to
intersecting subsets. The width of the thickening should be
suitably tuned, to be big enough to capture global properties, but
not too big in order to `identify' correctly the neighborhoods and
not to get to trivially intersecting cases (where all points are
in the neighborhood of all).In \cite{Major:2005fy} they proceed
using simplicial complexes and `nerves' to get the homology of the
anti-chain when considered as an approximation of a spacelike
surface of a manifold.

In both of these cases we get a cover of the anti-chain with
subsets corresponding to intersecting neighborhoods. A way to
define a topology on the anti-chain that would capture the spatial
topological properties of a manifold that would be the
approximation of the causet (giving fundamental status to the
causet), is the following. We consider the subsets and their
intersections and make a partial order of all these (neighborhoods
and intersections) where the order is set inclusion. Note that
this partial order is completely different that the causal partial
order of our initial causet. From this new partial order we may
consider the Alexandrov topology that would give us some `spatial
topology' on the spacelike surface in question. The Alexandrov
topology on a partial order, is defined to be the topology where
the open sets are the past sets of the partial order.

\begin{equation}\nonumber
S\subseteq X :\quad\forall\quad x,y\in X, x\in S \quad and\quad
y\preceq x \rightarrow y\in S
\end{equation}

Where $X$ is the partial order, and $S$ are the open subsets. To
calculate other properties, such as homology, we need to resort to
simplicial complexes as in \cite{Major:2005fy}.

In both cases, we may define a topology of a spatial surface of a
particular causet (if the causet is thought as a faithful
approximation of a continuous manifold). To get a `4-dimensional'
topology (\ie including the `time' dimension or else considering
Lorenzian rather than Euclidian manifold) we pack the slices
according to the parameter time.

 What
affects the result is

\begin{itemize}

\item The choice of the causet structure (in case it is
ambiguous).

\item The choice of slice (\ie the way we `foliated' the causet in
anti-chains).

\item The choice of the size of the `balls' used to define the
neighborhoods or the `thickness' of the slice, since those would
affect the spatial topology.

\end{itemize}

{\it In summa}, in the `algebraic approach', we have arrived at a
possible topology for our effective spacetime and, perhaps more
importantly, for doing this we have only used the set of different
histories.

\subsection{Comparing the approaches}

We shall now compare the two aforementioned ways of recovering
topology. Possible disagreements between them could stem from the
following:

\begin{itemize}

\item [(i)]The variation of the records was not `small enough', so
that the deduced topology does not correspond to the initial one,
which means that the first way failed.

\item[(ii)]The causet we chose is not the `real' one and one of
the ensuing ambiguities has possibly misled us, so that again the
first construction has failed. Note that had we considered all the
possible causets consistent with our data, we would find that one
of them agrees with the causet derived from the first way, unless
the first way failed due to the reason mentioned above.

\item[(iii)] Finally, the two ways of drawing the proximity
relation may intrinsically disagree, with the first way, thus
failing to identify the `real' nearest neighbor. This could also
indicate the incompleteness of our model of the experiment \eg
that the records we had, did not correspond to events.
\end{itemize}

These considerations rest on that from our causet reconstruction,
we had a unique unambiguous causet (up to a total time flip); or,
if more than one, that all resulted in homeomorphic topologies. In
the general case, it seems that we need to fix the interpretation
first. We can claim two things:

\begin{itemize}

\item[(i)] The state of the system is in superposition of
different topologies, one corresponding to each possible causet.
Further measurements that are made by varying the records will
result in a reduction of states. The probabilities for different
causets (amplitudes in the superpositions) could be recovered if
we repeated many times the whole procedure of varying the records.
If the procedure always yields a particular causet, we may
conclude that the state was in that `eigenstate' from the
beginning, and that it was us, that did not have access to the
records. Here, our failure to identify the correct state was due
to the fact that we were missing some information, namely, the
results of the measurements associated with the variation of the
records. We could therefore conclude that this failure was
basically an epistemic one, due to some kind of `classical
indeterminacy' see also \cite{case I}.

\item[(ii)] The state of the system is the equivalence class of
different causets, with equivalence relation being the possession
of same causal curves. In this context we cannot talk about the
topology of the equivalence class if there exist non-homeomorphic
topologies in the same equivalence class.

\end{itemize}

\subsection{Further Discussion}\label{stransprob}

If we take the point of view that the causet structure derived
from the set of histories is the best description we have for the
system and give ontological status to all the possible different
causets, we may as well enquire whether further measurements from
us may `reduce' the state to one particular causet (or to some
equivalence class thereof). We would then be talking of a
superposition of different causets. A way to handle these
`quantized causets' is by considering their incidence algebras as
in Raptis \cite{Raptis:1999ev}. It should be emphasized that if
the variations are indeed small, then the derived causet should
just be one of the possible causets derived without the variation.

The aforementioned method of determining the proximity by varying
the records could be viewed as a set of extra measurements that
restrict the class of possible causets---in a sense, as
determining the `actual' causet. Different procedures of deriving
the proximity will of course `favor' different causets.
Furthermore, allowing to vary the records lifts most ambiguities,
including the one involving an overall flip . If all procedures of
obtaining the proximity result to the same causet, we may as well
say that the structure of the effective spacetime of the system
was determined, but it was us that did not have access to the
records.

Since we are talking about an actual physical system in an actual
lab, one would not expect that the system actually `experiences'
such topological transitions. We could though imagine the
following Schr\"odinger's cat type of gedanken experiment. We have
a box with a particle inside. It is separated into two pieces, and
whether the wall between them falls or not depends on some
spin-half particle that is in a state of superposition. If we do
not have access to that particle, and we repeat many times the
experiment, getting both topologies is an actual possibility (pun
intended). Trying to determine which of the two is the `correct'
one would be equivalent to measuring the spin-half particle, and
that would then give a definite answer.

Finally, the other way of taking `seriously' all the possible
causets is to consider the physical states as being equivalence
classes of causets, related to a particular set of records (or
else, to a particular state of the record space). Then, small
record variations would merely pick one representative of our
physical state. Further investigation is needed to establish what
happens if bigger variations are allowed that would move from one
equivalence class to another. What remains to be considered, in
this case, is which of these classes of causets are close to
which. To do so, we would still have to define a notion of small
variation that would give us relative frequencies (however, the
variations should be big enough to move us out of the equivalence
class we happen to be, and into another).

\section{Conclusions}

Let us summarize what we have done. We have a laboratory in which
we explore a physical system whose effective spacetime, the
`arena' it lies, is unknown. We are able to run the experiments
sufficiently many times, either by leaving the initial conditions
unchanged, or by varying them. We also have another physical
system, whose configuration space is coined {\dff record space}.
In particular, we require from the record space to capture the
`spatio-temporal' properties of the system. We therefore have a
set of records for events .

We made the assumption that these records correspond to spacetime
points.\footnote{More precisely, `{\em effective}' spacetime
points, since we cannot have directly access to the `{\em real}'
spacetime, if this thing exists.} The assumption we made (about
the records) will then help us organize the information we have in
the present, as something that was a history. We will be dealing
with, in some sense,  a `timeless' theory, since time would emerge
merely as a better way to organize present data. To paraphrase
Wheeler (in his delayed choice experiments)
\cite{wheeler},\footnote{``{\em No phenomenon is a phenomenon
unless it is an observed phenomenon}''.} `events' become events
when some records are observed and when, by our (`delayed')
choice, those records are identified as events.

After multiple runs, we have a set of protocols (data-sheets).
Each protocol tells us which events occurred within a particular
experiment but it does not tells us in which order the events
occurred. This set of events is referred to as a history which, in
this context, correspond merely to a coarse grained `trajectory'
or else to a chain in a partial order that corresponds to the
effective spacetime. When the initial conditions remain unchanged,
the arising set of histories is treated as a decohering set.

More precisely, as a result of our observations, we have histories
and, in addition, their relative frequencies. This primary set of
histories we call {\dff fiducial set}. This corresponds to both
the sets of all possible events $\causet$ and of all possible
histories $\history$.

Using the fact that there are restrictions to the set of possible
histories due to causality, we proceed in section \ref{semergcaus}
to obtain a partially ordered set corresponding to the spacetime
(causal set). Due to our operationalistic methods, the causet is
recovered up to certain ambiguities (section \ref{Ambiguities in
Causet}).

Here we should point out that the above procedure accounts for the
following mathematical task.  Deriving a partial order given the
set of possible chains as mere subsets of the total partial order,
\ie without having the order of points in each of these chains.
This derivation is not unique and in section \ref{Ambiguities in
Causet} we classified the possible ambiguities.

We then considered two ways of recovering topology. The first one
involves some extra measurements, namely, varying the records to
get proximity and is similar to the one consider in the previous
paper \cite{case I}.

The second relies completely on the derived causet. We referred to
some work other work \cite{Bri-Gre91,Major:2005fy} on how to get
topology from a given causet in section \ref{salgrectop}. One
could use any scheme for deriving topology from a causet, and the
rest of the discussion would remain the same.

Since the above construction does not in general conclude to a
unique causet, we need some `interpretation' of the derived causet
before we can compare the different ways of deriving topology. We
may treat the possible causets as belonging to one equivalence
class and consider this equivalence class as a physical state.
This would run into problems if we would like to talk about
topology, if the algebraically derived topologies for different
causets in the same equivalence class are not homeomorphic.

Another way to interpret the set of all consistent with our
histories causets, is to assume that the system we consider
`lives' in a superposition of different effective spacetimes,
where each of the terms in the superposition,may have different
topology. In this case, we would consider the variation of the
records done in the `statistical' way of recovering the topology,
as a set of further measurements, causing the state to `reduce' to
a particular spacetime with its associated topology.

As a final point we would like to emphasize once again that we
recover histories \emph{operationalistically}. The record space is
\emph{the only} source of information we possess about the system
we explore. The effective topology is then regarded as the `best
possible' (:as realistic, or as pragmatic a) picture of the actual
background spacetime of the system in focus as one can acquire
from her `experimental intercourse' with it.

\paragraph{Acknowledgments.} We are grateful  to Jonathan Halliwell and Chris Isham for reading earlier
drafts and making many useful comments.We also thank Fay Dowker
and David Rideout for several comments about topology on a causet
and bringing to our attention reference \cite{Major:2005fy}. IR
acknowledges assistance from the EC in the form of a Marie Curie
European Reintegration Grant (ERG-CT-505432). PW acknowledges
partial funding from the Leventis Foundation. RRZ acknowledges the
financial support from the research grant No. 04-06-80215a from
RFFI. (Russian Basic Research Foundation). In the revised version
of the paper the authors express their gratitude to the referee of
\emph{International Journal of Theoretical Physics} for essential
remarks.


\begin{thebibliography}{99}

\bibitem{case I}
I.~Raptis, P.~Wallden, R.R.~Zapatrin, Spacetime topology from the
tomographic histories approach, Part I: Non-Relativistic Case, to
appear in IJTP volume proceedings of Glafka, gr-qc/0506088


\bibitem{EOT}       Barbour J 1999 \textit{The End of Time} (London: Weidenfeld and Nicolson; New York: Oxford University
                    Press)

\bibitem{Bri-Gre91}
G.~Brightwell and R.~Gregory, The Structure of Random Discrete
SpaceTime, \emph{Physical Review Letters,} \textbf{66}, 260--263
(1991)

\bibitem{Isham:1993qs}
  C.~J.~Isham,
  J.\ Math.\ Phys.\  {\bf 35} (1994) 2157
  [arXiv:gr-qc/9308006].


\bibitem{GH90c}
M.~Gell-{M}ann and J.~Hartle,
\newblock Alternative decohering histories in quantum mechanics.
\newblock In K. K. Phua and Y.~Yamaguchi, editors, {\em Proceedings of
the 25th International Conference on High Energy Physics,
Singapore, August, 2--8, 1990}, World Scientific, Singapore
(1990).

\bibitem{GH90a}
M.~Gell-{M}ann and J.~Hartle,
\newblock Quantum mechanics in the light of quantum cosmology.
\newblock In S.~Kobayashi, H.~Ezawa, Y.~Murayama, and S.~Nomura,
editors, {\em Proceedings of the Third International Symposium on
the Foundations of Quantum Mechanics in the Light of New
Technology}, pages 321--343. Physical Society of Japan, Tokyo
(1990).

\bibitem{GH90b}
M.~Gell-{M}ann and J.~Hartle,
\newblock Quantum mechanics in the light of quantum cosmology.
\newblock In W.~\.{Z}urek, editor, {\em Complexity, Entropy and the Physics of
Information, SFI Studies in the Science of Complexity, {Vol.
VIII}}, pages 425--458. Addison-Wesley, Reading (1990).

\bibitem{GH92}
M.~Gell-{M}ann and J.~Hartle,
\newblock Classical equations for quantum systems.
\newblock (1992)
\newblock UCSB preprint UCSBTH-91-15.

\bibitem{Gri84}
R. B. Griffiths,
\newblock Consistent histories and the interpretation of quantum mechanics.
\newblock {\em Journal of Statistical Physics}, {\bf 36}, 219
(1984).

\bibitem{Hal99}
\newblock J.~Halliwell,
\newblock Somewhere in the Universe: Where
is the information stored.
\newblock {\em Physical Review}, {\bf D60}, 105031 (1999).

\bibitem{Har91a}
J.~Hartle.
\newblock The quantum mechanics of cosmology,
\newblock In S.~Coleman, P.~Hartle, T.~Piran, and S.~Weinberg, editors, {\em
  Quantum Cosmology and Baby Universes}, World Scientific,
  Singapore (1991).

\bibitem{Har91b}
J.~Hartle.
\newblock Spacetime grainings in nonrelativistic quantum
mechanics,
\newblock {\em Physical Revew}, {\bf D44}, 3173 (1991).

\bibitem{Har93a}
J.~Hartle,
\newblock Spacetime quantum mechanics and the quantum mechanics of spacetime.
\newblock In {\em Proceedings on the 1992 Les Houches School, Gravitation and
  Quantisation} (1993).



 \bibitem{Kribs:2005yc}
  D.~W.~Kribs and F.~Markopoulou,
  arXiv:gr-qc/0510052.


\bibitem{Lloyd:2005js}
  S.~Lloyd,
  arXiv:quant-ph/0501135

\bibitem{Major:2005fy}
  S.~Major, D.~Rideout and S.~Surya,
  Spatial Hypersurfaces in Causal Set Cosmology,
eprint gr-qc/0506133.



  \bibitem{Omn88a}
\newblock R.~Omn\`es,
\newblock Logical reformulation of quantum mechanics, {I.} {F}oundations.
\newblock {\em Journal of Statistical Physics}, {\bf 53}, 893 (1988).

\bibitem{Omn88b}
\newblock R.~Omn\`es.
\newblock Logical reformulation of quantum mechanics. {II.} {I}nterferences and
  the {E}instein-{P}odolsky-{R}osen experiment.
\newblock {\em Journal of Statistical Physics}, {\bf 53}, 933 (1988).

\bibitem{Omn88c}
\newblock R.~Omn\`es.
\newblock Logical reformulation of quantum mechanics. {III.} Classical limit
  and irreversibility.
\newblock {\em Journal of Statistical Physics}, {\bf 53}, 957 (1988).

\bibitem{Omn89}
\newblock R.~Omn\`es,
\newblock Logical reformulation of quantum mechanics. {IV. CTRL} {P}rojectors in
  semiclassical physics,
\newblock {\em Journal of Statistical Physics}, {\bf 57}, 357 (1989).

\bibitem{Omn90}
\newblock R.~Omn\`es,
\newblock From Hilbert space to common sense: {A} synthesis of recent
  progress in the interpretation of quantum mechanics,
\newblock {\em Annals of Physics (NY)}, {\bf 201}, 354 (1990).

\bibitem{Omn92}
\newblock R.~Omn\`es,
\newblock Consistent interpretations of quantum mechanics.
\newblock {\em Reviews of Modern Physics}, {\bf 64}, 339 (1992).



\bibitem{rbr}
\newblock R.~Breslav,
\newblock Private communication

\bibitem{Raptis:1999ev}
  I.~Raptis, Algebraic Quantization of Causal Sets,
  Int.\ J.\ Theor.\ Phys.\  {\bf 39} (2000) 1233;
eprint gr-qc/9906103



\bibitem{Sor91a}
R. D. Sorkin,
\newblock Finitary substitute for continuous topology.
\newblock {\em Int. J. Theor. Phys.}, {\bf 30} (1991) 923--947

\bibitem{Sor91b}
R. D. Sorkin,
\newblock Spacetime and causal sets.
\newblock In J. C. D'Olivo, E.~Nahmad-Achar, M.~Rosenbaum, M. P. Ryan, L. F. Urrutia, and
F.~Zertuche, editors, {\em Relativity and Gravitation: Classical
  and Quantum}, pages 150--173. World Scientific, Singapore, 1991.


\bibitem{wheeler} J. A. Wheeler, \newblock It from Bit.
\newblock In W. H. Zurek, editor, {\em Complexity, entropy, and the physics of information: the proceedings of the
1988 Workshop on Complexity, Entropy, and the Physics of
Information}, Santa-Fe Institute, New Mexico, Addison-Wesley,
1990.


  \bibitem{prg}
R.R.Zapatrin, Pre-Regge calculus: topology via logic,
  Int.\ J.\ Theor.\ Phys.\  {\bf 32} 779--799 (1993)

\end{thebibliography}
 \end{document}